\documentclass[namedreferences]{solarphysics}

\usepackage[hyperref,optionalrh]{spr-sola-addons} 
\usepackage{graphicx}        
\usepackage{amssymb}        
\usepackage{color}           
\usepackage{breakurl}        
\renewcommand{\vec}[1]{{\mathbfit #1}}

\renewcommand{\div}{ \mathrm{div} }
\newcommand{\cur}{ \mathrm{curl} }

\newcommand{\pder}[2]{ \frac{\partial #1}{\partial #2} }
\newcommand{\grad}{ {\bf \nabla } }

\newcommand{\solphys}{{\it Sol. Phys.}}

\chardef\us=`\_

\begin{document}

\begin{article}
\begin{opening}

\title{Thermal Trigger for Solar Flares II: Effect of the
Guide Magnetic Field}

\author[addressref={1},corref,email={leonid.ledentsov@gmail.com}]{\inits{L.S.}\fnm{Leonid}~\lnm{Ledentsov}
\orcid{0000-0002-2701-8871}}
\address[id={1}]{Sternberg Astronomical Institute, 
Moscow State University, 
Moscow 119234, Universitetsky pr., 13, Russia}

\runningauthor{Ledentsov L.S.}
\runningtitle{Thermal Trigger for Solar Flares II}

\begin{abstract}
We investigate the effect of the thermal imbalance on the structural stability of the magnetohydrodynamic model 
of the preflare current layer (\citeauthor{2021SoPh..296...74L}, \solphys{} \textbf{296}, 74, 
\citeyear{2021SoPh..296...74L}). 
The piecewise homogeneous model of the current layer is supplemented 
by a magnetic field longitudinal with respect to the direction of the current.
It is shown that the presence of a weak longitudinal field 
does not change the previously calculated spatial period of the thermal instability  
in the most expected range of the parameters of the preflare current layer 
and, moreover, contributes to the formation of the instability. 
On the other hand, a strong longitudinal magnetic field contributes to the spatial stabilization of the current layer.
\end{abstract}
\end{opening}

\section{Introduction}
     \label{sec1} 

The evolution of the coronal active region associated with a solar flare contains two main stages. 
The first stage consists in the slow accumulation of the free energy of the magnetic field 
in the vicinity of the preflare current layer \citep{1971JETP...33..933S}. 
The second stage corresponds to the rapid transformation of the accumulated magnetic energy 
into thermal and kinetic energy of the plasma, 
the energy of accelerated particles and electromagnetic radiation \citep{2011LRSP....8....6S}. 
The first stage proceeds mainly with the preservation of the global topology of the magnetic field, 
but during the second stage the topology changes dramatically \citep{2012A&A...538A.138O}. 
A redistribution of magnetic fluxes occurs. 
Magnetic reconnection process is a key mechanism 
for changing the topology of the magnetic field \citep{2002A&ARv..10..313P}.

Ideal conditions for magnetic reconnection occur at X-type zero lines of a magnetic field. 
But the reconnection process under the actual conditions in the solar atmosphere 
is released at the separator which differs from the X-type neutral line 
in that the separator has a longitudinal magnetic field \citep{1988SvA....32..308G, 1988SoPh..117...77G}. 
The longitudinal field, often called the guide field, at the separator decreases the reconnection rate 
primarily by the fact that it changes the balance of forces in the reconnecting current layer. 
The pressure of the surrounding plasma and the external magnetic field now restrains 
not only the pressure of the current layer plasma, 
but also the pressure of the longitudinal magnetic field \citep{2006ASSL..341.....S}.
If the longitudinal field could be effectively accumulated inside the current layer, 
its pressure would impose strong limitations on the layer compression and, hence, on the rate of reconnection. 

In a real plasma, dissipative effects associated with the longitudinal field are important. 
As soon as the value of the longitudinal field inside the layer becomes comparable 
to the main field (the field whose lines are reconnected in the current layer), 
the gradient drift of charged particles forms a circular current in the plane perpendicular to the main current in the layer 
\citep{2006ASSL..340.....S}. 
The Ohmic dissipation of the current, circulating around the layer, 
gives rise to an outward diffusion of the longitudinal field from the current layer and to the Joule heating of the plasma. 
It is worth noting that the full flux of the longitudinal field is preserved, and the plasma is heated 
due to the energy of the reconnected fluxes of the main field. 
Due to the dissipation of the circular current, the plasma penetrates relatively easily through the longitudinal magnetic field. 
This process limits the accumulation of the longitudinal magnetic field in the layer. 

It turns out that on the one hand, the longitudinal magnetic field decreases the rate of plasma inflow into the layer. 
On the other hand, if the plasma compression inside the layer is small, 
the contribution of the longitudinal field to the balance of pressure and Joule heating of the current layer 
can be insignificant as long as its value does not exceed the value of the main field. 
The model of the super-hot turbulent-current layer describing the impulsive phase of a solar flare 
shows a weak compression of the plasma \citep{1985SoPh...95..141S}.
During a flare, the longitudinal field may not have a significant effect on the properties 
of the current layer in a wide range of conditions 
\citep{1985SoPh..102...79S}.
A different situation can arise in the forming preflare current layer.
Direct evidence of the accumulation of the longitudinal field 
inside the current layer is obtained on the basis of magnetic measurements in laboratory experiments
\citep{2009PhLA..373.1460F}. 
The value of the ring current during the formation of the current layer turns out 
to be comparable to the total current in the layer. 
Thus, dissipative effects can be weak
and the longitudinal field can reach large values
in the preflare current layer.

In the first article of this series \citep{2021SoPh..296...74L}, a simple magnetohydrodynamic (MHD) model 
of the preflare current layer was proposed to analyze the effect of thermal imbalance 
on structural stability of the current layer. 
The problem of small perturbations in such a layer has several unstable solutions, 
the most important of which under the conditions of the solar atmosphere 
is a thermal instability \citep{1965ApJ...142..531F}. 
The instability in the linear phase grows with the characteristic time of radiative plasma cooling. 
The fragmentation of the current layer across the direction of the current 
with a spatial period of about $1-10$ Mm is the result of the instability.

In this article, the model of the preflare current layer is supplemented with a longitudinal magnetic field 
in order to study its influence on the formation of a thermal instability. 
The structure of the article is as follows. 
The problem of small perturbations in a current layer with a longitudinal magnetic field is solved in Section~\ref{sec2}. The effect of viscosity is discussed in Section~\ref{sec3}. 
The solutions found are applied to the conditions of the preflare coronal plasma in Section~\ref{sec4}.
Conclusions are given in Section~\ref{sec5}.

\section{Non-Neutral Current Layer without Viscosity
}
\label{sec2}

In this section, the piecewise homogeneous model of the current layer 
presented in \cite{2021SoPh..296...74L} will be generalized by adding a magnetic field longitudinal 
with respect to the direction of the current (Figure \ref{fig1}).
The model is considered in the magnetohydrodynamic (MHD) approximation.
An infinite current layer located in the ($x,z$) plane has a half-thickness $a$.
The magnetic field inside the layer has only one component $B_z$ directed along the current.
Outside the current layer, the magnetic field has a second component $B_0=100 {\rm \,G}$ directed 
against the $x$-axis for positive $y$ and along the $x$-axis for negative $y$.
Constant by piece magnetic field components will allow us 
to find an analytical solution to the problem of small perturbations 
of a non-neutral current layer, taking into account thermal, dissipative, and compressibility effects.
The spatial uniformity of the magnetic field requires that the electric currents 
flow over the surfaces of the current layer and have zero thickness.
The currents are directed along the $z$-axis and screen the component of the magnetic field $B_0$, 
but freely pass the component $B_z$ into the current layer.
Thus, the half-thickness of the current layer in the considered model 
is a free parameter, which determines the region of space free from counter-directional magnetic fluxes.

Focusing on the evolution of the current layer structure along the direction of the current, 
we will neglect the dependence of the perturbations on the $x$-coordinate.
This means that in this article we do not consider the well-known tearing instability \citep{1963PhFl....6..459F}
and the associated filamentation of the current layer into separate current bundles.
The considered structural instability, in contrast to the tearing instability, 
has a thermal nature and leads to a periodic perturbation of the layer along the current.
Its growth rate is independent of the current layer thickness
(see Equation 33 in \citeauthor{2021SoPh..296...74L}, \citeyear{2021SoPh..296...74L}).
However, the half-thickness of the current layer affects the spatial scale of the instability.
Earlier, we identified two approximations for the spatial scale, 
corresponding to thin and thick current layers with respect to the perturbation length 
(Equations 35 and 37 in \citeauthor{2021SoPh..296...74L}, \citeyear{2021SoPh..296...74L}, respectively).
In this article, we will consider two realistic values of the current layer half-thickness 
($a=10^5 {\rm \, cm}$ and $a=10^6 {\rm \, cm}$, 
see Chapter 8 in \citeauthor{2006ASSL..341.....S}, \citeyear{2006ASSL..341.....S})
to study the effect of the current layer thickness on the simulation results.

The surrounding plasma has a coronal temperature $T_0=10^6 {\rm\,K}$ 
and an increased concentration $n_0=10^{10} {\rm\,cm}^{-3}$
caused by the raking of the frozen-in plasma during the formation of the preflare current layer.
The plasma is assumed to be ideal outside the current layer, 
while dissipative effects are important inside it: 
the finite electrical $\sigma$ and thermal $\kappa$ conductivity of the plasma, 
as well as its radiative cooling $\lambda$.
The radiative cooling function $\lambda \,(n,T)=n^2L(T)$ contains 
the total radiative loss rate $L(T)$ that is calculated 
from the CHIANTI atomic database \citep{2019ApJS..241...22D} 
for an optically thin medium with coronal abundance of elements (see Figure 1 in \cite{2021SoPh..296...74L}).
For clarity, the effects of viscosity are omitted in this section and considered separately in Section~\ref{sec3}.
In this case, a sufficient set of MHD equations is as follows
\citep{1958ForPh...6..437S}:
\begin{figure}
\begin{center}
\vspace{3mm}
\includegraphics*[width=0.6\linewidth]{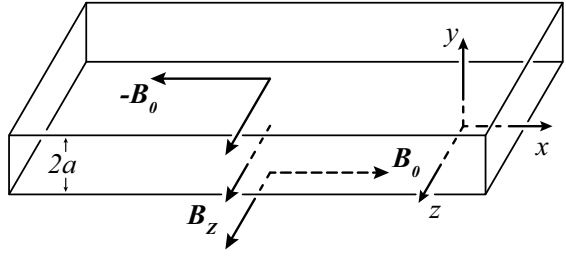}
\end{center}
\caption{Location of the current layer in the coordinate system.}
\label{fig1}
\end{figure}
\begin{displaymath}
    \frac{\partial n}{\partial t} + \div \, (n \vec{v})
    = 0 \, ,
\end{displaymath}
\begin{displaymath}
    \mu n \, \frac{\rm{d} \vec{v}}{{\rm d} t}
  = - \grad (2 n k_{_{ \rm B }} T)
  - \frac{1}{4\pi} \, ( \vec{B} \times \cur \vec{B} ) \, ,
\end{displaymath}
\begin{displaymath}
    \frac{2 n k_{_{ \rm B }} }{\gamma -1} \,
    \frac{{\rm d} T}{{\rm d} t}
  - 2 k_{_{ \rm B }} T \, \frac{ {\rm d} n }{ {\rm d} t } \\
  = \frac{ c^2 }{ ( 4 \pi )^2 \sigma } \, ( {\cur} {\vec B} )^2
  + {\div} \, ( \kappa \grad T ) - \lambda \, (n,T) \, ,
  \end{displaymath}
\begin{displaymath}
    \frac{\partial {\vec B}}{\partial t}
  = {\cur} \, ( {\vec v} \times {\vec B} )
  - \frac{ c^2 }{ 4 \pi } \, {\cur}
    \left( \frac{1}{\sigma} \, {\cur} {\vec B} \right) ,
\end{displaymath}
\begin{equation}
    {\div} {\vec B} = 0 \, .
\label{01}
\end{equation}
Here, $ \mu =1.44 \, m_H $, $ m_H $ is the mass of the hydrogen atom,
$ k_{_{ \rm B }} $ is the Boltzmann constant.
The heat capacity ratio is assumed $\gamma = 5/3$ for simplicity. 
$T$ is the temperature, $n$ is the plasma density, $v$ is the plasma velocity, and $B$ is the magnetic field. 
Anisotropy of thermal and electrical conductivity in the presence of a magnetic field inside the current layer
is an essential distinguishing feature of this model in comparison with the \cite{2021SoPh..296...74L} model.
However, the method for solving the problem remains the same. 
The solution of the system for small perturbations will be found 
separately outside the layer and inside the layer. 
Then the found solutions will be sewn on the boundary, 
which is a tangential MHD discontinuity.
The plasma concentration $n_s$ and temperature $T_s$ inside the current layer 
will serve as free parameters of the model.

\subsection{Outside the Current Layer}
\label{sec2.1}

The plasma is at rest $v_0=0$, and the dissipative effects are negligible 
$\sigma \to \infty$, $\kappa=0$, $\lambda=0$ outside the current layer.
The symmetry of the model allows us to consider only the positive $y$ axis.
The solution to the problem of small perturbations is assumed 
to be periodic in the direction of the current and to decay with distance from the current layer:
\begin{displaymath}
    f(y,z,t)
  = f_0
  + f_1(y) \, {\rm exp} \, ( - i \omega t + i k_z z ) \, ,
\end{displaymath}
\begin{displaymath}
    f_1(y)=f_1 \, {\rm exp} \, [ - k_{y1} (y-a) ] \, , 
\end{displaymath}
where perturbation amplitudes are
\begin{displaymath}
    f_1 \equiv \{{v_{x1}, v_{y1}, v_{z1}, n_1, T_1, B_{x1}, B_{y1}, B_{z1}}\} \, ,
\end{displaymath}

The linearized system of Equations \ref{01} takes the form:
\begin{equation}
    i \omega \, n_1
  = - k_{y1} \, n_0 v_{y1} + i k_z \, n_0 v_{z1} \, ,
    \label{02}
\end{equation}
\begin{equation}
    i \omega \, \mu n_0 v_{x1}
  = i k_{z} \, \frac{ B_z  }{ 4 \pi }\,B_{x1}  \, ,
    \label{03}
\end{equation}
\begin{equation}
    i \omega \, \mu n_0 v_{y1}
  = - k_{y1} \, 2k_{_{ \rm B }}  (n_0 T_1 + T_0 n_1)
    + k_{y1} \, \frac{ B_0  }{ 4 \pi }\,B_{x1}  \, ,
    \label{04}
\end{equation}
\begin{equation}
    i \omega \, \mu n_0 v_{z1}
  = i k_{z} \, 2k_{_{ \rm B }}  (n_0 T_1 + T_0 n_1)
  - i k_{z} \, \frac{ B_0  }{ 4 \pi } \, B_{x1} \, ,
    \label{05}
\end{equation}
\begin{equation}
    ( \gamma - 1 ) \, T_0 n_1 = n_0 T_1 \, ,
    \label{06}
\end{equation}
\begin{equation}
    i \omega \, B_{x1}
  = k_{y1} \, B_0 \, v_{y1} - i k_z \, B_0 \, v_{z1} - i k_z \, B_z \, v_{x1} \, ,
    \label{07}
\end{equation}
\begin{equation}
    i \omega \, B_{y1}
  = - i k_z \, B_z \, v_{y1} \, ,
    \label{08}
\end{equation}
\begin{equation}
    i \omega \, B_{z1}
  = - k_{y1} \, B_z \, v_{y1} \, .
    \label{09}
\end{equation}
The determinant of a homogeneous system of linear Equations \ref{02}--\ref{09}
must be equal to zero for a nontrivial solution to exist.
Therefore, the dispersion relation 
\begin{equation}
    k_{y1}^2 = 
    \left(\frac{1}{ \varepsilon_S + 
    \varepsilon_A (1+\frac{n_s}{n_0}\frac{1}{\varepsilon})^{-1}} - \frac{1}{\varepsilon}\right) 
    \frac{{\it \Gamma}^2}{V_Z^2}\, ,
    \label{10}
\end{equation}
must be fulfilled outside the current layer.
The speed of sound and $x$-axis projection of the Alfv\'en speed outside the current layer, 
the phase velocity of the perturbation, and the Alfv\'en speed inside the current layer:
\begin{equation}
    V_S = \sqrt{ \frac{ 2 \gamma k_{_{ \rm B }} T_0 }{ \mu }} \, ,
    \qquad
        V_A = \frac{B_0}{ \sqrt{ 4 \pi n_0 \mu } } \, ,
        \qquad
        V = \frac{\omega}{k_z} \, ,
        \qquad
        V_{Z} = \frac{B_z}{ \sqrt{ 4 \pi n_s \mu } } \, 
    \label{20}
\end{equation}
are included in dimensionless quantities
\begin{equation}
    \varepsilon_S=\frac{V_S^2}{V_Z^2} \, ,
    \qquad
    \varepsilon_A=\frac{V_A^2}{V_Z^2} \, ,
    \qquad
    \varepsilon=\frac{V^2}{V_Z^2} \, .
    \label{20}
\end{equation}
${\it \Gamma} = - i \omega$ is the instability increment.
%
%

\vspace{1mm}

%
%
\subsection{Inside the Current Layer}
\label{sec2.2}

The plasma is also at rest $v_s=0$, but the dissipative effects must be considered inside the current layer.
The solution is also sought in the form of a sum of a constant term and a perturbation
\begin{displaymath}
    f(y,z,t) = f_s
             + f_2(y) \, {\rm exp} \, (-i\omega t + ik_zz) \, .
\end{displaymath}
Following \cite{2021SoPh..296...74L}, we consider the dependence of the perturbation on the coordinate $y$ 
in the form of a hyperbolic sine for odd perturbations in $y$ 
\begin{displaymath}
\left\{
\begin{array}{c}
v_{x2}(y)\\
v_{y2}(y)\\
B_{x2}(y)\\
B_{y2}(y)
\end{array}
\right\} =
\left\{
\begin{array}{c}
v_{x2}\\
v_{y2}\\
B_{x2}\\
B_{y2}
\end{array}
\right\}
{\rm sinh} \, (k_{y2} y) \, ,
\end{displaymath}
and a hyperbolic cosine for even perturbations in $y$
\begin{displaymath}
\left\{
\begin{array}{c}
v_{z2}(y)\\
n_2(y)\\
T_2(y)\\
B_{z2}(y)
\end{array}
\right\} =
\left\{
\begin{array}{c}
v_{z2}\\
n_2\\
T_2\\
B_{z2}
\end{array}
\right\}
{\rm cosh} \, (k_{y2} y) \, .
\end{displaymath}
%
%

Inside the current layer, the linearized system of Equations \ref{01} takes the form:
\begin{equation}
    i \omega \, n_{2}
  = k_{y2} \, n_s v_{y2} + i k_z \, n_s v_{z2} \, ,
    \label{13}
\end{equation}
\begin{equation}
    i \omega \, \mu n_s v_{x2}
  = i k_{z} \, \frac{ B_z  }{ 4 \pi }\,B_{x2}  \, ,
    \label{14}
\end{equation}
\begin{equation}
    i \omega \, \mu n_s v_{y2}
  = k_{y2} \, 2 k_{_{ \rm B }} ( n_s T_2 + T_s n_2 ) \, ,
    \label{15}
\end{equation}
\begin{equation}
    i \omega \, \mu n_s v_{z2}
  = i k_{z} \, 2 k_{_{ \rm B }} ( n_s T_2 + T_s n_2 ) \, ,
    \label{16}
\end{equation}
\begin{equation}
    i \omega \, \frac{ 2 k_{_{ \rm B }} n_s }{ \gamma - 1} \, T_2
  - i \omega \, 2 k_{_{ \rm B }} T_s \, n_2 =  ( \kappa_{\|} k_z^2 - \kappa_{\bot} k_{y2}^2 ) \, T_2
   + \pder{ \lambda }{ T } \, T_2 + \pder{ \lambda }{ n } \, n_2 \, ,
    \label{17}
\end{equation}
\begin{equation}
    i \omega \, B_{x2}
  =  - i k_z \, B_z \, v_{x2} + (\nu_{m\bot} k_z^2 - \nu_{m\|}k_{y2}^2) \, B_{x2} \, ,
    \label{18}
\end{equation}
\begin{equation}
    i \omega \, B_{y2}
  =  - i k_z \, B_z \, v_{y2} + \nu_{m\bot}( i k_z k_{y2} \, B_{z2} + k_{z}^2 \, B_{y2} ) \, ,
    \label{19}
\end{equation}
\begin{equation}
    i \omega \, B_{z2}
  =  k_{y2} \, B_z \, v_{y2} + \nu_{m\bot}( i k_z k_{y2} \, B_{y2} - k_{y2}^2 \, B_{z2} )  \, ,
    \label{20}
\end{equation}
\begin{equation}
 k_{y2} \, B_{y2} = - i k_{z} \, B_{z2} \, .
    \label{21}
\end{equation}
Hereinafter, the magnetic viscosity along and across the magnetic field is denoted as
\begin{displaymath}
    \nu_{m\|}
  =  \frac{ c^2 }{4 \pi \sigma_\|} \, ,
  \qquad
    \nu_{m\bot}
  =  \frac{ c^2 }{4 \pi \sigma_\bot} \,  .
\end{displaymath}
The set of Equations \ref{13}--\ref{21} splits into two sets and, as a consequence, has two dispersion relations at once.
Equations \ref{14} and \ref{18} contain only $v_{x2}$ and $B_{x2}$ perturbations and 
give the dispersion relation
\begin{equation}
    k_{y2}^2 = 
    \left(\frac{1+\varepsilon^{-1}}{ {\it \Gamma} \tau_\sigma \, \upsilon \varepsilon } - \frac{\varsigma}{\varepsilon}\right)
    \frac{{\it \Gamma}^2}{V_Z^2} \, .
    \label{22}
\end{equation}
This equation contains the degree of anisotropy of the electrical conductivity 
$\varsigma=\sigma_\|/\sigma_\bot$ and the dimensionless quantity
\begin{displaymath}
    \upsilon = \frac{ 8 \pi k_{_{ \rm B }} T_s n_s }{ B_z^2 \, \varepsilon } \, .
\end{displaymath}

The second dispersion relation follows from Equations \ref{13}, \ref{15}--\ref{17}, and \ref{19}--\ref{21}.
Expressing the difference $k_z^2 - k_{y2}^2$ from Equation \ref{22}, 
we can exclude it from the second dispersion relation and
and find the equation for the instability growth rate (${\it \Gamma}=-i\omega$)
\begin{eqnarray}
{\it \Gamma}^{\,4} \,
  &-& \left[ \, 
\frac{ 1 }
         { 1+\delta }\,{\it \Gamma}_1 \,+\,
     \frac{ 1+\upsilon(\varsigma-1)(1+\delta) }{ 1+\upsilon(\varsigma-1) }\,{\it \Gamma}_3 \,
        \right]
    {\it \Gamma}^{\,3}
    \nonumber \\&+& \left[ \,          {\it \Gamma}_1 {\it \Gamma}_3 \,+\,
     \frac{ (- \alpha(\beta - \alpha)^{-1}+\upsilon(\varsigma-1))(1+\delta) }
     { 1+\upsilon(\varsigma-1) }\,{\it \Gamma}_2{\it \Gamma}_3 \,
        \right]
 {\it \Gamma}^{\,2}
         \nonumber \\&-& {\it \Gamma}_1{\it \Gamma}_2{\it \Gamma}_3\,{\it \Gamma} = 0 \, ,
    \label{23}
\end{eqnarray}
where the approximate roots of Equation \ref{23} are
\begin{eqnarray}
    {\it \Gamma}_0&=&0 \, ,    
    \qquad\qquad\qquad\qquad {\it \Gamma}_1=
    \frac{ ( 1+\delta )(1+\varepsilon^{-1}) }{ \tau_\sigma \, (1+\upsilon(\varsigma-1)) } \, ,    
    \nonumber \\
    {\it \Gamma}_2 &=&
    \frac{ \beta-\alpha }{ \tau_\lambda } \, \frac{ \delta }{ 1+\delta } \, ,
    \qquad \quad \, {\it \Gamma}_3 =
    \frac{ 1 }{ \tau_\kappa \, \upsilon(\varkappa-\varsigma) \, \delta } \, .
    \label{24}
\end{eqnarray}
The roots ${\it \Gamma}_0, {\it \Gamma}_1, {\it \Gamma}_2$ repeat the roots in the neutral current layer \citep{2021SoPh..296...74L}
with some corrections that disappear at $B_z=0$.
The root ${\it \Gamma}_3$ is new and depends on the degrees of anisotropy 
of the thermal conductivity $\varkappa=\kappa_\|/\kappa_\bot$ 
and electrical conductivity $\varsigma=\sigma_\|/\sigma_\bot$.
Equations \ref{24} are the solution to Equation \ref{23} for ${\it \Gamma}_2 \ll {\it \Gamma}_1 \ll {\it \Gamma}_3$ 
and the values of the fractions in square brackets of Equation \ref{23} are close to one.
The weaker the longitudinal magnetic field and the anisotropy caused by it, the better are these conditions.

The dimensionless parameter 
\begin{equation}
\delta = 
         \left({ \frac{1}{\gamma-1} - \frac{\tau_\kappa}{\tau_\sigma} \left(1+\frac{1}{\varepsilon}\right)}\right)^{-1}
    \label{25}
\end{equation}
differs from the one entered in \cite{2021SoPh..296...74L} by the factor $1+\varepsilon^{-1}$. 
The characteristic times 
\begin{equation}
    \tau_\sigma 
  =  \frac{ \mu \, \nu_{m\|}}{2k_{_{ \rm B }} T_s} \, ,
  \qquad
\tau_\kappa
  = \frac{ \mu \, \kappa_\bot }{ (2k_{_{ \rm B }})^2 T_s n_s } \, ,
\qquad
    \tau_\lambda
  =  \frac{ 2 k_{_{ \rm B }} T_s n_s }{ \lambda } \, 
    \label{26}
\end{equation}
correspond to those in \cite{2021SoPh..296...74L}, taking into account the anisotropy of the transfer coefficients.

%
%

\vspace{1mm}

%
%
\subsection{Boundary of the Current Layer}
\label{sec2.3}

The tangential discontinuity is located at the boundary of the current layer.
Zero plasma velocity ($v_0=0, v_s=0$) and the absence of a component of the magnetic field 
normal to the discontinuity surface indicate this \citep{2015PhyU...58..107L}.

The boundary condition for the tangential discontinuity is that 
the total gas and magnetic pressures on both sides of the discontinuity are equal \citep{1956TrFIAN...13..64S}.
In addition, velocity perturbations $v_x, v_y, v_z$ will lead 
to a wave-like curvature of the discontinuity surface \citep{2021SoPh..296...74L}.
The linearized boundary conditions are written as follows:
\begin{equation}
    n_0 T_1 + T_0 n_1 - \frac{ B_0 B_{x1} }{ 8 \pi k_{_{ \rm B }} } + \frac{ B_z B_{z1} }{ 8 \pi k_{_{ \rm B }} }
  = ( n_s T_2 + T_s n_2 + \frac{ B_z B_{z2} }{ 8 \pi k_{_{ \rm B }} } ) \, {\rm cosh} \, ( k_{y2} a ) \, .
    \label{27}
\end{equation}
\begin{equation}
    v_{y1} = \pm \, v_{y2} \, {\rm sinh} \, ( k_{y2} a ) \, .
    \label{28}
\end{equation}
The first three terms on the left side of Equation \ref{27} can be expressed in terms of perturbation $v_{y1}$
using Equation \ref{04}, and the last one using Equation \ref{09}.
The first two terms on the right side of Equation \ref{26} can be expressed in terms of perturbation $v_{y2}$
using Equation \ref{15}, and the last one using Equations \ref{20}--\ref{20}.
Then Equation \ref{27} takes the form
\begin{equation}
    -  \left(\frac{n_0}{n_s} + \frac{k_{y1}^2 V_Z^2}{{\it \Gamma}^2} \right) \frac{v_{y1}}{k_{y1}} \,
  =  \left( 1+\frac{k_{y2}^2 V_Z^2}{{\it \Gamma}^2+ {\it \Gamma} \nu_{m\bot}(k_z^2-k_{y2}^2)} \right) 
  \frac{v_{y2}}{k_{y2}} \, {\rm cosh} \, ( k_{y2} a ) \, .
    \label{29}
\end{equation}

An additional dispersion relation is obtained after dividing Equation \ref{28} by Equation \ref{29}
\begin{equation}
    \pm \, \frac{k_{y1}}{ \frac{n_0}{n_s} + \frac{k_{y1}^2 V_Z^2}{{\it \Gamma}^2} } = 
    \frac{k_{y2}\, {\rm tanh} \, ( k_{y2} a )}{ 1+\frac{k_{y2}^2 V_Z^2}{{\it \Gamma}^2+ {\it \Gamma} \nu_{m\bot}(k_z^2-k_{y2}^2)}} \, .
    \label{30}
\end{equation}
Wave numbers $k_{y1}$ and $k_{y2}$ can be eliminated from Equation \ref{30} by using
Equations \ref{10} and \ref{22}, respectively
\begin{equation}
    \frac{\frac{1}{ \varepsilon_S + 
    \varepsilon_A (1+\frac{n_s}{n_0}\frac{1}{\varepsilon})^{-1}} - \frac{1}{\varepsilon} }
    { \left( \frac{n_0}{n_s} + \frac{1}{ \varepsilon_S + 
    \varepsilon_A (1+\frac{n_s}{n_0}\frac{1}{\varepsilon})^{-1}} - \frac{1}{\varepsilon} \right)^2 }
    =
    \frac{\left(\frac{1+\varepsilon^{-1}}{ {\it \Gamma} \tau_\sigma \, \upsilon \varepsilon } - \frac{\varsigma}{\varepsilon}\right)
    {\rm tanh}^2
    \left[ \frac{a{\it \Gamma}}{V_Z}
    \left(\frac{1+\varepsilon^{-1}}{ {\it \Gamma} \tau_\sigma \, \upsilon \varepsilon }-\frac{\varsigma}{\varepsilon}\right)^{1/2}
    \right]}
    { \left( 1- \frac{\frac{1+\varepsilon^{-1}}{ {\it \Gamma} \tau_\sigma \, \upsilon \varepsilon } - \frac{\varsigma}{\varepsilon}}
    {\varsigma(1+\varepsilon^{-1}) - 1 - \varsigma {\it \Gamma} \tau_\sigma \, \upsilon (\varsigma-1)}\right)^2}
     \, .
    \label{31}
\end{equation}
Equation \ref{31} allows, knowing the growth increment (Equations \ref{24}),
determine the spatial period of the
instability 
\begin{equation}
    l = \frac{2 \pi}{ k_z }= \frac{2 \pi V_Z \sqrt{ -\varepsilon }}{\it \Gamma} \, .
    \label{32}
\end{equation}

\section{Non-Neutral Current Layer with Viscosity
}
\label{sec3}

This section will consider the effect of viscosity on the stability of the current layer model.
The system of Equations \ref{01} will include new terms
\citep{2006ASSL..340.....S}:
\begin{displaymath}
    \frac{\partial n}{\partial t} + \div \, (n \vec{v})
    = 0 \, ,
\end{displaymath}
\begin{displaymath}
    \mu n \, \frac{\rm{d} \vec{v}}{{\rm d} t}
  = - \grad (2 n k_{_{ \rm B }} T)
  - \frac{1}{4\pi} \, ( \vec{B} \times \cur \vec{B} )+ \pder{\sigma_{\alpha \beta}}{r_\beta}  \, ,
\end{displaymath}
\begin{displaymath}
    \frac{2 n k_{_{ \rm B }} }{\gamma -1} \,
    \frac{{\rm d} T}{{\rm d} t}
  - 2 k_{_{ \rm B }} T \, \frac{ {\rm d} n }{ {\rm d} t } \\
  = \frac{ c^2 }{ ( 4 \pi )^2 \sigma } \, ( {\cur} {\vec B} )^2 + \pder{}{r_\alpha} (\sigma_{\alpha \beta} v_\beta)
  + {\div} \, ( \kappa \grad T ) - \lambda \, (n,T) \, ,
  \end{displaymath}
\begin{displaymath}
    \frac{\partial {\vec B}}{\partial t}
  = {\cur} \, ( {\vec v} \times {\vec B} )
  - \frac{ c^2 }{ 4 \pi } \, {\cur}
    \left( \frac{1}{\sigma} \, {\cur} {\vec B} \right) ,
\end{displaymath}
\begin{equation}
    {\div} {\vec B} = 0 \, .
\label{33}
\end{equation}
Here $\alpha$ and $\beta$ are tensor indices corresponding to the spatial coordinates $x$, $y$, and $z$.
The viscous stress tensor $\sigma_{\alpha \beta}$ will have a main influence on the equation of motion, 
because when linearizing the energy conservation equation, 
it will give a second-order term with respect to the perturbed quantities.
For simplicity, we will neglect all viscosity coefficients except $\eta_0$ \citep{1965RvPP....1..205B}.
This is true for a sufficiently strong magnetic field.
Then the viscous stress tensor will contain three components
\begin{displaymath}
    \sigma_{xx} = \frac{\eta_0}{2}  (W_{xx}+W_{yy}) \, , \qquad
    \sigma_{yy} = \frac{\eta_0}{2}  (W_{xx}+W_{yy}) \, , \qquad 
    \sigma_{zz} = \eta_0 W_{zz} \, ,
\end{displaymath}
Where the strain-rate tensor has the form
\begin{displaymath}
    W_{\alpha \beta}=\pder{v_\alpha}{r_\beta}+\pder{v_\beta}{r_\alpha}-\frac{2}{3}\delta_{\alpha\beta}\,\div\,\vec{v} \, .
\end{displaymath}

Dissipative effects are irrelevant outside the current layer, including viscosity effects.
Therefore, all conclusions drawn in Section~\ref{sec2.1} remain valid for the viscous case.
Let's move on to considering the interior of the current layer.

%
%
\subsection{Inside the Current Layer}
\label{sec3.1}

The solution inside the current layer is sought in the same form as in Section~\ref{sec2.2}
The system of Equations \ref{33} is linearized with regard to viscosity as follows:

\begin{equation}
    i \omega \, n_{2}
  = k_{y2} \, n_s v_{y2} + i k_z \, n_s v_{z2} \, ,
    \label{34}
\end{equation}
\begin{equation}
    i \omega \, \mu n_s v_{x2}
  = i k_{z} \, \frac{ B_z  }{ 4 \pi }\,B_{x2}  \, ,
    \label{35}
\end{equation}
\begin{equation}
    i \omega \, \mu n_s v_{y2}
  = k_{y2} \, 2 k_{_{ \rm B }} ( n_s T_2 + T_s n_2 ) - k_{y2}^2 \, 
  \eta_0 v_{y2} + \frac{2}{3} i \omega k_{y2} \frac{\eta_0}{n_s} n_2 \, ,
    \label{36}
\end{equation}
\begin{equation}
    i \omega \, \mu n_s v_{z2}
  = i k_{z} \, 2 k_{_{ \rm B }} ( n_s T_2 + T_s n_2 ) + 2 k_{z}^2 \, 
  \eta_0 v_{z2} + \frac{2}{3} i \omega i k_{z} \frac{\eta_0}{n_s} n_2 \, ,
    \label{37}
\end{equation}
\begin{equation}
    i \omega \, \frac{ 2 k_{_{ \rm B }} n_s }{ \gamma - 1} \, T_2
  - i \omega \, 2 k_{_{ \rm B }} T_s \, n_2 =  ( \kappa_{\|} k_z^2 - \kappa_{\bot} k_{y2}^2 ) \, T_2
   + \pder{ \lambda }{ T } \, T_2 + \pder{ \lambda }{ n } \, n_2 \, ,
    \label{38}
\end{equation}
\begin{equation}
    i \omega \, B_{x2}
  =  - i k_z \, B_z \, v_{x2} + (\nu_{m\bot} k_z^2 - \nu_{m\|}k_{y2}^2) \, B_{x2} \, ,
    \label{39}
\end{equation}
\begin{equation}
    i \omega \, B_{y2}
  =  - i k_z \, B_z \, v_{y2} + \nu_{m\bot}( i k_z k_{y2} \, B_{z2} + k_{z}^2 \, B_{y2} ) \, ,
    \label{40}
\end{equation}
\begin{equation}
    i \omega \, B_{z2}
  =  k_{y2} \, B_z \, v_{y2} + \nu_{m\bot}( i k_z k_{y2} \, B_{y2} - k_{y2}^2 \, B_{z2} )  \, ,
    \label{41}
\end{equation}
\begin{equation}
 k_{y2} \, B_{y2} = - i k_{z} \, B_{z2} \, .
    \label{42}
\end{equation}
As we can see, the system of Equations \ref{34}--\ref{42} also splits into two sets, 
as the system of Equations \ref{13}--\ref{21}, 
and Equation \ref{22} remains valid.
However, the equation for the instability growth rate will change to
\begin{eqnarray}
{\it \Gamma}^{4}
  &-& \left[ 
  \frac{ 9-\frac{1}{\upsilon} }{ (9\upsilon\varsigma+4-\varsigma)(1+\delta) }{\it \Gamma}_1 +
  \frac{ 9\upsilon\varsigma(1+\delta)+4-\varsigma }{ 9\upsilon\varsigma+4-\varsigma }{\it \Gamma}_3 +
  (1+\upsilon(\varsigma-1)){\it \Gamma}_4  \right]{\it \Gamma}^{3}
\nonumber \\&+& \left[ 
   \frac{ 9(1+\delta)-\frac{1}{\upsilon} }{ (9\upsilon\varsigma+4-\varsigma)(1+\delta) }{\it \Gamma}_1{\it \Gamma}_3 +
   \frac{ 1 }{ 1+\delta }{\it \Gamma}_1{\it \Gamma}_4 \right. \nonumber \\&+& \left.
   \frac{ (9\upsilon\varsigma+\frac{ - \alpha}{\beta - \alpha}(4-\varsigma))(1+\delta) }
   { 9\upsilon\varsigma+4-\varsigma }{\it \Gamma}_2{\it \Gamma}_3 +
   (1+\upsilon(\varsigma-1)(1+\delta)){\it \Gamma}_3{\it \Gamma}_4
\right] {\it \Gamma}^{2}
\nonumber \\&-& \left[ 
   \frac{ 9-\frac{ - \alpha}{\beta - \alpha}\frac{1}{\upsilon} }
   { 9\upsilon\varsigma+4-\varsigma }{\it \Gamma}_1{\it \Gamma}_2{\it \Gamma}_3 + 
   {\it \Gamma}_1{\it \Gamma}_3{\it \Gamma}_4 + 
   \left(\frac{ - \alpha}{\beta - \alpha}+\upsilon(\varsigma-1)\right)(1+\delta){\it \Gamma}_2{\it \Gamma}_3{\it \Gamma}_4
         \right] {\it \Gamma}
\nonumber \\&+&   
   {\it \Gamma}_1{\it \Gamma}_2{\it \Gamma}_3{\it \Gamma}_4 =0 \, .
    \label{43}
\end{eqnarray}
The approximate roots
\begin{eqnarray}
    {\it \Gamma}_1 &=&
    \frac{ (1+\delta)(1+\varepsilon^{-1}) }{ \tau_\sigma } \, ,
    \quad \quad {\it \Gamma}_2 =
    \frac{ \beta - \alpha }{ \tau_\lambda } \, \frac{\delta}{1+\delta} \, ,
    \nonumber \\
    {\it \Gamma}_3 &=&
    \frac{1}{ \tau_\kappa \, \upsilon(\varkappa-\varsigma) \, \delta} \, ,
    \quad \qquad \,\,\, {\it \Gamma}_4 =
    \frac{ 3 }{ \tau_\eta \, \upsilon(9\upsilon\varsigma+4-\varsigma) } \, 
    \label{44}
\end{eqnarray}
are the solution to Equation \ref{43} for ${\it \Gamma}_2 \ll {\it \Gamma}_1 \sim {\it \Gamma}_4 \ll {\it \Gamma}_3$ 
and coefficients before the roots in Equation \ref{43} are close to one. 
The new root ${\it \Gamma}_4$ contains another characteristic time
\begin{displaymath}
  \tau_\eta
  = \frac{ \eta_0 }{ 2k_{_{ \rm B }} T_s n_s } \, .
\end{displaymath}

%
%

\vspace{1mm}

%
%
\subsection{Boundary of the Current Layer}
\label{sec3.2}

Viscosity does not allow directly expressing the combination $n_s T_2 + T_s n_2$ 
in terms of the velocity perturbation $v_{y2}$ from Equation \ref{36}.
It is required to additionally involve Equations \ref{34} and \ref{37}.
Therefore, an additional term appears on the right side of Equation \ref{31} due to the viscosity
\begin{equation}
    \frac{\frac{1}{ \varepsilon_S + 
    \varepsilon_A (1+\frac{n_s}{n_0}\frac{1}{\varepsilon})^{-1}} - \frac{1}{\varepsilon} }
    { \left( \frac{n_0}{n_s} + \frac{1}{ \varepsilon_S + 
    \varepsilon_A (1+\frac{n_s}{n_0}\frac{1}{\varepsilon})^{-1}} - \frac{1}{\varepsilon} \right)^{\!2} } =
    \frac{\left(\frac{1+\varepsilon^{-1}}{ {\it \Gamma} \tau_\sigma \, \upsilon \varepsilon } - \frac{\varsigma}{\varepsilon}\right)
    {\rm tanh}^2
    \left[ \frac{a{\it \Gamma}}{V_Z}
    \left(\frac{1+\varepsilon^{-1}}{ {\it \Gamma} \tau_\sigma \, \upsilon \varepsilon }-\frac{\varsigma}{\varepsilon}\right)^{1/2}
    \right]}
    { \left( \frac{2}{3} + \frac{1}{3}\,\frac{\varsigma {\it \Gamma} \tau_\eta \upsilon + \theta}
    {1\,-\, 2\,{\it \Gamma} \tau_\eta \upsilon} - 
    \frac{\frac{1+\varepsilon^{-1}}{ {\it \Gamma} \tau_\sigma \, \upsilon \varepsilon } - \frac{\varsigma}{\varepsilon}}
    {\varsigma(1+\varepsilon^{-1}) - 1 - \varsigma {\it \Gamma} \tau_\sigma \, \upsilon (\varsigma-1)}\right)^{\!2}}
     \, .
    \label{45}
\end{equation}

\section{Instability of the Preflare Current Layer}
\label{sec4}

The root ${\it \Gamma}_2$ from Equations \ref{24} and \ref{44}
is the only one that corresponds to the MHD approximation 
in the absence of a magnetic field in the current layer \citep{2021SoPh..296...74L}. 
Large values of the growth rate ${\it \Gamma}$ lead to very small spatial scales of the instability. 
Thus, it has been shown that the approximate analytical roots ${\it \Gamma}_1$ and ${\it \Gamma}_2$
(see Equation 13 in \citeauthor{2021SoPh..296...74L}, \citeyear{2021SoPh..296...74L})
are in very good agreement with the exact solution of the equation for the growth increment
(see Equation 12 in \citeauthor{2021SoPh..296...74L}, \citeyear{2021SoPh..296...74L}). 
It is also shown that ${\it \Gamma}_1 \gg {\it \Gamma}_2$ in the conditions of the solar corona. 
New approximate roots in Equations \ref{24} and \ref{44} found in Sections~\ref{sec2.2} and \ref{sec3.1} respectively
obey the conditions ${\it \Gamma}_3 \gg {\it \Gamma}_1$ and ${\it \Gamma}_3 \sim {\it \Gamma}_1$ 
at least for a weak $B_z$ component of the magnetic field. 
This means that the instability associated with them also does not satisfy the considered MHD approximation. Therefore, in this section, only the root ${\it \Gamma}_2$ is considered, 
the appearance of which does not undergo changes 
when a longitudinal field with or without viscosity is added to the model of the preflare current layer. 
However, the parameter $\delta$ contains changes (see Equation \ref{25}). 
First, an additional factor $1+\varepsilon^{-1} < 1$ for positive ${\it \Gamma}$ appears in its denominator. 
Secondly, the characteristic time $\tau_\kappa$ contained in it is calculated 
not through the longitudinal coefficient of thermal conductivity, but through the transverse one (see Equation \ref{26}). 
Both of these factors decrease the value $\frac{\tau_\kappa}{\tau_\sigma} \left(1+\frac{1}{\varepsilon}\right)$, 
and therefore contribute to the formation of the instability, 
since the instability occurs at ${\it \Gamma}>0$ 
or at $\frac{\tau_\kappa}{\tau_\sigma} \left(1+\frac{1}{\varepsilon}\right)<\frac{1}{\gamma-1}$. 
For further calculations, it will be assumed that, for some reason, the thermal instability has come into play. 
That is, in fact, we will assume that (see Equation 33 in \citeauthor{2021SoPh..296...74L}, \citeyear{2021SoPh..296...74L})
\begin{equation}
   {\it \Gamma} =
    \frac{ 2 }{ 5 } \, \frac{ \beta-\alpha }{ \tau_\lambda } \,.
    \label{46}
\end{equation}

The dispersion relation in Equation \ref{45} differs from that in Equation \ref{31}
only in the additional term in the denominator of the right-hand side, 
associated with the presence of viscosity. 
Considering that the denominator also includes a term that depends on plasma conductivity, 
and that it will vary in calculations, then we assume $\eta_0=0$, 
implying in the calculations the general effects of viscosity and electrical conductivity. 
Also, for definiteness, we set $\varsigma=2$ \citep{1965RvPP....1..205B}.

Dispersion relations in Equations \ref{31} and \ref{45} are more complex than 
a similar relation in the absence of a magnetic field 
(see Equation 32 in \citeauthor{2021SoPh..296...74L}, \citeyear{2021SoPh..296...74L}). 
An iterative method is used to find a solution. 
Under the conditions of an active region in the solar corona, as a first approximation, 
we can neglect the terms with the order $\varepsilon^{-1}$ on the right-hand side of the dispersion relations, 
as was done in the search for thin and thick approximations 
(Equations 34--37 in \citeauthor{2021SoPh..296...74L}, \citeyear{2021SoPh..296...74L}), 
as well as the term $\frac{n_s}{n_0}\frac{1}{\varepsilon}$ 
on the left-hand side of the dispersion relations in Equations \ref{31} and \ref{45}.
Moreover, if we take into account the formation of the current layer, 
then the longitudinal field outside the layer should be $\frac{n_s}{n_0}$
times less than the field inside the layer due to the freezing-in effect. 
Then the terms $\frac{n_s}{n_0}\frac{1}{\varepsilon}$ 
on the left side of the dispersion relations in Equations \ref{31} and \ref{45}
will change to $\frac{n_0}{n_s}\frac{1}{\varepsilon}$. 
In any case, these terms will not change the final result due to their smallness.
Then a quadratic equation is formed with respect to $\varepsilon^{-1}$
\begin{eqnarray}
    \left(\frac{1}{\varepsilon}\right)^2&+&
    \left[\frac{({\it \Gamma} \tau_\sigma \, \upsilon \varepsilon-(\varsigma-1)^{-1})^2}
    { {\it \Gamma} \tau_\sigma \, \upsilon \varepsilon \,
    {\rm tanh}^2
    \left[ \frac{a{\it \Gamma}}{V_Z}
    ({\it \Gamma} \tau_\sigma \, \upsilon \varepsilon)^{-1/2} \right] } \,-\,
    2\left( \frac{n_0}{n_s} + \frac{1}{ \varepsilon_S + 
    \varepsilon_A  } \right)
    \right] \, \frac{1}{\varepsilon}     \nonumber
 \\
    &+& \left( \frac{n_0}{n_s} + \frac{1}{ \varepsilon_S + 
    \varepsilon_A  } \right)^2 \,-\, 
    \frac{1}{ \varepsilon_S + 
    \varepsilon_A  } \,
    \frac{({\it \Gamma} \tau_\sigma \, \upsilon \varepsilon-(\varsigma-1)^{-1})^2}
    { {\it \Gamma} \tau_\sigma \, \upsilon \varepsilon \, 
    {\rm tanh}^2
    \left[ \frac{a{\it \Gamma}}{V_Z}
    ({\it \Gamma} \tau_\sigma \, \upsilon \varepsilon)^{-1/2} \right]} \,=\,0 \, .
\nonumber
\end{eqnarray}
The found $\varepsilon^{-1}$ is substituted into the original dispersion relation everywhere, 
except for the numerator of the right-hand side, and so on, 
until the difference between the new and the old $\varepsilon^{-1}$ is less than a certain threshold.
In these calculations, the threshold was taken to be $10^{-4}$.
Further, the found $\varepsilon^{-1}$ is recalculated to the spatial scale of instability
$l$ using Equation \ref{32}.

\begin{figure}
\vspace{2mm}
\begin{center}
\includegraphics*[width=\linewidth]{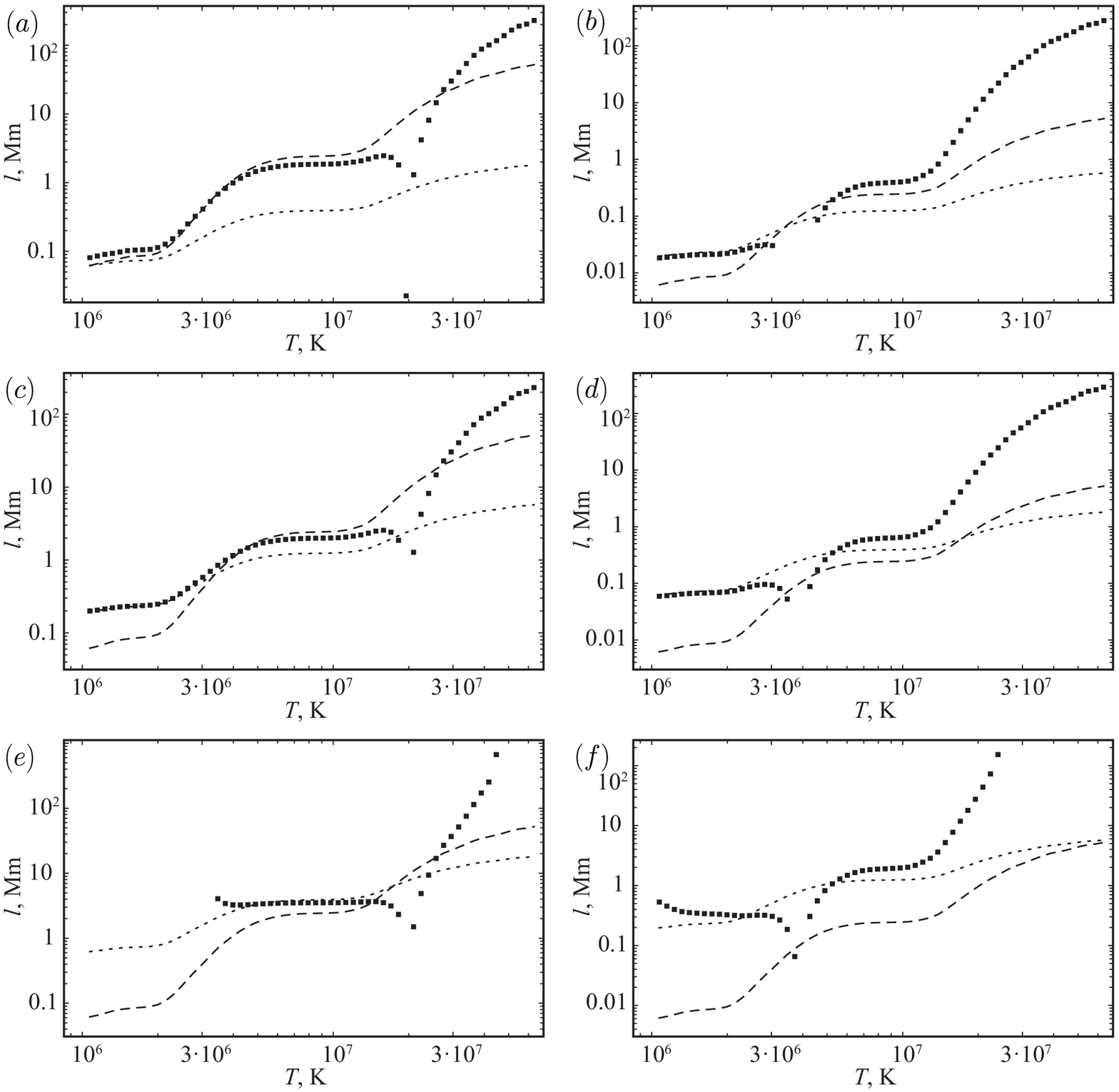}
\end{center}
\caption{The spatial period of the instability depending on the temperature of the current layer.
Parameters of the coronal plasma: $n_0=10^{10} {\rm \, cm}^{-3}$,
$a=10^5 {\rm \, cm}$, 
$B_0=100 {\rm \, G}$.
Parameters of the current layer: 
(a) $n_s=10^{11} {\rm \, cm}^{-3}$, $B_z=10^{-3} {\rm \, G}$, 
$\sigma_\|=10^{11} {\rm \, s}^{-1}$;
(b) $n_s=10^{11} {\rm \, cm}^{-3}$, $B_z=10^{-3} {\rm \, G}$, 
$\sigma_\|=10^{12} {\rm \, s}^{-1}$;
(c) $n_s=10^{12} {\rm \, cm}^{-3}$, $B_z=10^{-2} {\rm \, G}$, 
$\sigma_\|=10^{11} {\rm \, s}^{-1}$;
(d) $n_s=10^{12} {\rm \, cm}^{-3}$, $B_z=10^{-2} {\rm \, G}$, 
$\sigma_\|=10^{12} {\rm \, s}^{-1}$;
(e) $n_s=10^{13} {\rm \, cm}^{-3}$, $B_z=10^{-1} {\rm \, G}$, 
$\sigma_\|=10^{11} {\rm \, s}^{-1}$;
(f) $n_s=10^{13} {\rm \, cm}^{-3}$, $B_z=10^{-1} {\rm \, G}$, 
$\sigma_\|=10^{12} {\rm \, s}^{-1}$.
Squares are solutions to the dispersion relations in Equations \ref{31} and \ref{45}.
Dashed and dotted lines are thin and thick approximations, respectively, 
without taking into account the longitudinal magnetic field (Equations 35 and 37 in \citeauthor{2021SoPh..296...74L}, \citeyear{2021SoPh..296...74L}).
}
\label{fig2}
\end{figure}
%
\begin{figure}
\vspace{2mm}
\begin{center}
\includegraphics*[width=\linewidth]{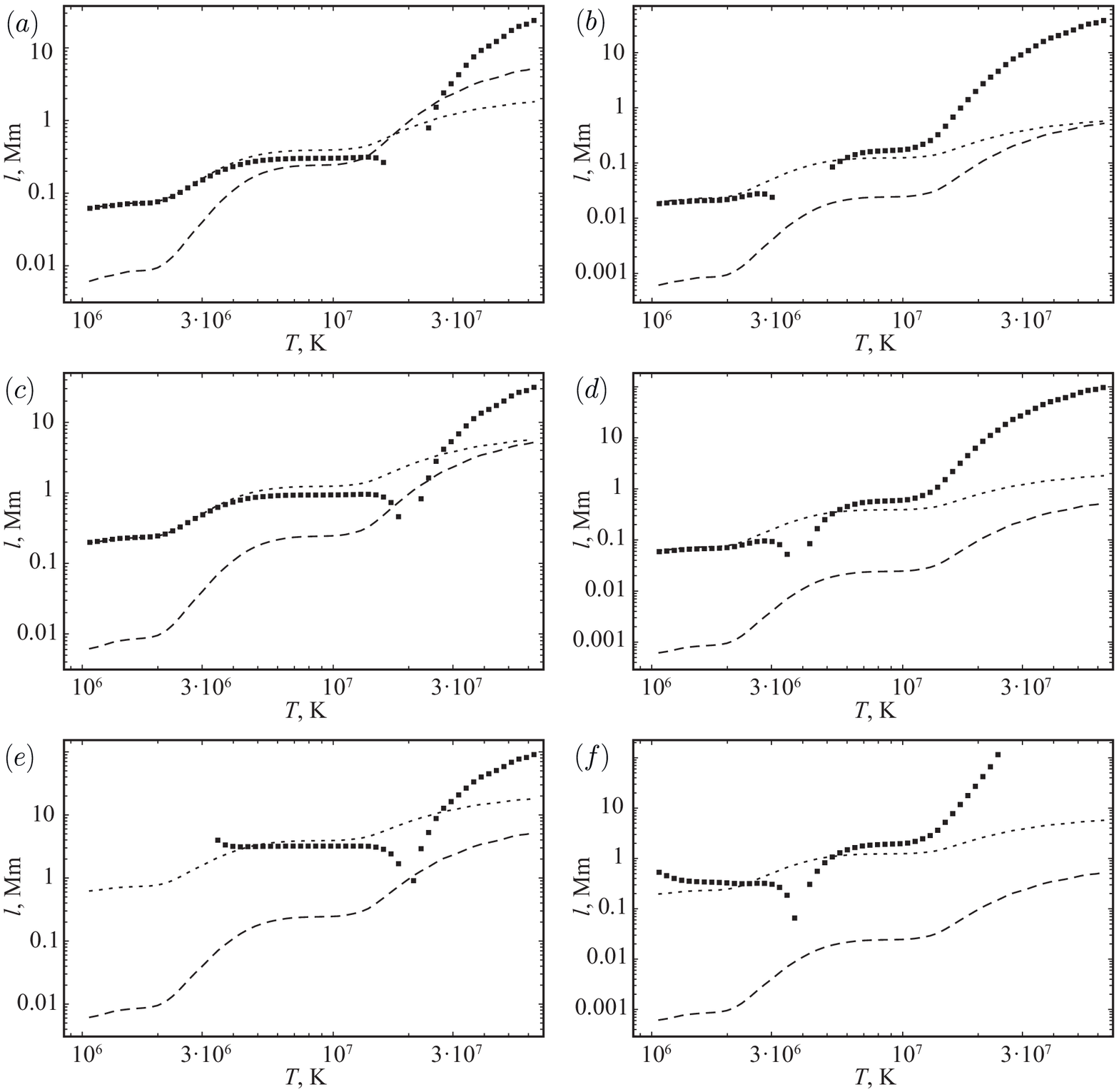}
\end{center}
\caption{The spatial period of instability depending on the temperature of the current layer.
Parameters of the coronal plasma: $n_0=10^{10} {\rm \, cm}^{-3}$,
$a=10^6 {\rm \, cm}$, 
$B_0=100 {\rm \, G}$.
Parameters of the current layer: 
(a) $n_s=10^{11} {\rm \, cm}^{-3}$, $B_z=10^{-3} {\rm \, G}$, 
$\sigma_\|=10^{11} {\rm \, s}^{-1}$;
(b) $n_s=10^{11} {\rm \, cm}^{-3}$, $B_z=10^{-3} {\rm \, G}$, 
$\sigma_\|=10^{12} {\rm \, s}^{-1}$;
(c) $n_s=10^{12} {\rm \, cm}^{-3}$, $B_z=10^{-2} {\rm \, G}$, 
$\sigma_\|=10^{11} {\rm \, s}^{-1}$;
(d) $n_s=10^{12} {\rm \, cm}^{-3}$, $B_z=10^{-2} {\rm \, G}$, 
$\sigma_\|=10^{12} {\rm \, s}^{-1}$;
(e) $n_s=10^{13} {\rm \, cm}^{-3}$, $B_z=10^{-1} {\rm \, G}$, 
$\sigma_\|=10^{11} {\rm \, s}^{-1}$;
(f) $n_s=10^{13} {\rm \, cm}^{-3}$, $B_z=10^{-1} {\rm \, G}$, 
$\sigma_\|=10^{12} {\rm \, s}^{-1}$.
Squares are solutions to the dispersion relations in Equations \ref{31} and \ref{45}.
Dashed and dotted lines are thin and thick approximations, respectively, 
without taking into account the longitudinal magnetic field (Equations 35 and 37 in \citeauthor{2021SoPh..296...74L}, \citeyear{2021SoPh..296...74L}).
}
\label{fig3}
\end{figure}
Figures \ref{fig2} and \ref{fig3} show the calculation results 
for the most interesting combinations of coronal plasma parameters. 
The calculations were performed for two values of the half-thickness of the current layer:
$a=10^5 {\rm \, cm}$ (Figure \ref{fig2}) and $a=10^6 {\rm \, cm}$ (Figure \ref{fig3}).
The first value is closer to the thin approximation of the current layer 
in the model without a guide magnetic field \citep{2021SoPh..296...74L}, 
while the second value is closer to the thick approximation.
For comparison, the calculation of the dispersion relations of our model (squares) 
are shown together with the thin and thick approximations from \cite{2021SoPh..296...74L} 
(dashed and dotted lines, respectively). 
The gaps in the squares correspond to the temperatures 
at which the declared calculation accuracy of $\varepsilon^{-1}$ was not achieved. 
Common to all plots is that the spatial scale of the instability does not differ much 
from that obtained without taking into account the longitudinal magnetic field \citep{2021SoPh..296...74L} 
in the temperature range of the preflare current layer $T_s=5\times10^6-10^7$ K
that is most interesting from the point of view of expectations and is $1-10$ Mm. 
At the same time, the model of a current layer with a longitudinal magnetic field
gives a steeper increase in the scale of the instability at temperatures $T_s>10^7$ K. 
Sharp dips at temperature $T_s \approx 2\times10^7$ in Figures \ref{fig2}a, c, e and \ref{fig3}a, c, e 
and temperature $T_s \approx 4\times10^6$ in Figures \ref{fig2}b, d, f and \ref{fig3}b, d, f 
are another interesting feature. 
The position of the dip is determined by the denominator 
of the right-hand side of the Equations \ref{31} and \ref{45}. 
Thus, the dip structure can depend not only on the effective conductivity of the current layer plasma, 
but also on its viscosity.

It should be noted that all calculations were carried out 
with a weak longitudinal magnetic field $B_z=10^{-3}-10^{-1}$ G. 
An increase in the longitudinal field leads to imaginary values of the wavenumber $k_z$, 
and hence to spatial stabilization of the instability on the same scales. 
Thus, we can say that a strong longitudinal magnetic field leads to stabilization 
of the considered model of a preflare current layer. 
In this context, it is worth mentioning the stabilizing effect of the longitudinal magnetic field 
on the tearing instability \citep{1963PhFl....6..459F, 1993SSRv...65..253S}, 
namely, the tearing instability of a compressible non-neutral current layer \citep{1993ARep...37..282V}. 
In any case, if the instability leads to an increase in the energy of the magnetic field, 
this is disadvantageous from the standpoint of the energetic principle of stability.
At the same time, a weak longitudinal field component contributes to the formation of instability, 
as discussed earlier. 
Finally, it should be noted that the longitudinal magnetic field 
is included in the expressions for the growth rate of instability in Equations \ref{24} and \ref{44}
and the dispersion relations in Equations \ref{31} and \ref{45} only as part of the square of the Alfv\'en velocity $V_Z^2$, 
which means that the properties of the considered instability 
do not depend on whether the field in the layer is directed along the current or against it.

\section{Conclusion}
\label{sec5}

The model of the preflare current layer \citep{2021SoPh..296...74L} 
is supplemented with a guide magnetic field. 
The magnetic field changes the model in several ways. 
First, the anisotropy of heat and electrical conductivity 
associated with the presence of a magnetic field 
diversifies the possible growth rates of instability in Equations \ref{24} and \ref{44}
and complicates the dispersion relations in Equations \ref{31} and \ref{45}. 
Despite this, the only physically significant growth rate for coronal plasma conditions remains 
the growth rate inversely proportional to the characteristic time of radiative plasma cooling (Equation \ref{46}), 
and the dispersion relations produce the spatial scale of the instability close to $1-10$ Mm
for the most acceptable temperature range of the preflare current layer $T_s=5\times10^6-10^7$ K. 
At the same time, an unusual sharp dip appears on the spatial scale profile, 
which can noticeably diversify the spatial periods of the instability 
admissible by the model towards smaller scales (Figures \ref{fig2} and \ref{fig3}).

Second, a strong longitudinal magnetic field leads to spatial stabilization of the instability, 
while a weak one, on the contrary, contributes to its formation. 
Both of these processes are important for the physics of solar flares. 
The first promotes the accumulation of free energy of the magnetic field 
in the vicinity of the preflare current layer, 
the second releases the accumulated energy in the form of a solar flare. 
Thus, adding a longitudinal magnetic field to the preflare current layer model 
does not change the leading role of the thermal instability in triggering solar flares
and enriches the model with an additional mechanism. 
It can be qualitatively described as follows. 
Active regions in real solar flares can have a very complex magnetic field structure. 
The longitudinal magnetic fields inevitably present in them stabilize the current layers, 
but the evolution of the active region at some point in time leads 
to a relatively weak longitudinal field, which opens up the possibility 
for the formation of the thermal instability of the preflare current layer with its further destruction 
and the beginning of the process of fast magnetic reconnection, that is, a flare.

\vspace{1cm}
\noindent{\bf Disclosure of Potential Conflicts of Interest} The author declares that there are no conflicts of interest.

\bibliographystyle{spr-mp-sola}
\bibliography{sola_bibliography}  

\begin{thebibliography}{21}
\ifx\bisbn     \undefined \def\bisbn  #1{ISBN #1}\fi
\ifx\binits    \undefined \def\binits#1{#1}\fi
\ifx\bauthor   \undefined \def\bauthor#1{#1}\fi
\ifx\batitle   \undefined \def\batitle#1{#1}\fi
\ifx\bjtitle   \undefined \def\bjtitle#1{\textit{#1}}\fi
\ifx\bvolume   \undefined \def\bvolume#1{\textbf{#1}}\fi
\ifx\byear     \undefined \def\byear#1{#1}\fi
\ifx\bissue    \undefined \def\bissue#1{#1}\fi
\ifx\bfpage    \undefined \def\bfpage#1{#1}\fi
\ifx\blpage    \undefined \def\blpage #1{#1}\fi
\ifx\burl      \undefined \def\burl#1{#1}\fi
\ifx\href      \undefined \def\href#1#2{#2}\fi
\ifx\betal     \undefined \def\betal{et al.}\fi
\ifx\bctitle   \undefined \def\bctitle#1{#1}\fi
\ifx\beditor   \undefined \def\beditor#1{#1}\fi
\ifx\bbtitle   \undefined \def\bbtitle#1{\textit{#1}}\fi
\ifx\bedition  \undefined \def\bedition#1{#1}\fi
\ifx\bseriesno \undefined \def\bseriesno#1{\textbf{#1}}\fi
\ifx\blocation \undefined \def\blocation#1{#1}\fi
\ifx\bsertitle \undefined \def\bsertitle#1{\textit{#1}}\fi
\ifx\bsnm      \undefined \def\bsnm#1{#1}\fi
\ifx\bsuffix   \undefined \def\bsuffix#1{#1}\fi
\ifx\bparticle \undefined \def\bparticle#1{#1}\fi
\ifx\barticle  \undefined \def\barticle#1{}\fi
\ifx\binstitute  \undefined \def\binstitute#1{#1}\fi
\ifx\bpublisher  \undefined \def\bpublisher#1{#1}\fi
\ifx\doiurl    \undefined \def\doiurl#1{\href{#1}{DOI}}\fi
\makeatletter
\def\safeHref#1#2#3{\in@{http}{#2}\ifin@\href{#2}{#3}\else\href{#1#2}{#3}\fi}
\makeatother
\ifx\adsurl    \undefined
  \def\adsurl#1{\safeHref{https://ui.adsabs.harvard.edu/abs/}{#1}{ADS}}\fi
\ifx\arxivurl  \undefined
  \def\arxivurl#1{\safeHref{http://arxiv.org/abs/}{#1}{arXiv}}\fi
\ifx\botherref \undefined \def\botherref#1{}\fi
\ifx\url       \undefined \def\url#1{#1}\fi
\ifx\bchapter  \undefined \def\bchapter#1{}\fi
\ifx\bbook     \undefined \def\bbook#1{}\fi
\ifx\bcomment  \undefined \def\bcomment#1{#1}\fi
\ifx\oauthor   \undefined \def\oauthor#1{#1}\fi
\ifx\citeauthoryear \undefined\def \citeauthoryear#1{#1}\fi
\def\endbibitem {}
\ifx\bconflocation  \undefined \def\bconflocation#1{#1} \fi

\bibitem[\protect\citeauthoryear{{Braginskii}}{1965}]{1965RvPP....1..205B}
\begin{barticle}
\bauthor{\bsnm{{Braginskii}}, \binits{S.I.}}:
\byear{1965},
\batitle{{Transport Processes in a Plasma}}.
\bjtitle{Rev. Plasma Phys.}
\bvolume{1},
\bfpage{205}.
\adsurl{1965RvPP....1..205B}.
\end{barticle}
\endbibitem

\bibitem[\protect\citeauthoryear{{Dere} et~al.}{2019}]{2019ApJS..241...22D}
\begin{barticle}
\bauthor{\bsnm{{Dere}}, \binits{K.P.}},
\bauthor{\bsnm{{Del Zanna}}, \binits{G.}},
\bauthor{\bsnm{{Young}}, \binits{P.R.}},
\bauthor{\bsnm{{Landi}}, \binits{E.}},
\bauthor{\bsnm{{Sutherland}}, \binits{R.S.}}:
\byear{2019},
\batitle{{CHIANTI{\textemdash}An Atomic Database for Emission Lines. XV.
  Version 9, Improvements for the X-Ray Satellite Lines}}.
\bjtitle{\apjs}
\bvolume{241},
\bfpage{22}.
\doiurl{https://doi.org/10.3847/1538-4365/ab05cf}.
\adsurl{2019ApJS..241...22D}.
\end{barticle}
\endbibitem

\bibitem[\protect\citeauthoryear{{Field}}{1965}]{1965ApJ...142..531F}
\begin{barticle}
\bauthor{\bsnm{{Field}}, \binits{G.B.}}:
\byear{1965},
\batitle{{Thermal Instability.}}
\bjtitle{\apj}
\bvolume{142},
\bfpage{531}.
\doiurl{https://doi.org/10.1086/148317}.
\adsurl{1965ApJ...142..531F}.
\end{barticle}
\endbibitem

\bibitem[\protect\citeauthoryear{{Frank}, {Bugrov}, and
  {Markov}}{2009}]{2009PhLA..373.1460F}
\begin{barticle}
\bauthor{\bsnm{{Frank}}, \binits{A.}},
\bauthor{\bsnm{{Bugrov}}, \binits{S.}},
\bauthor{\bsnm{{Markov}}, \binits{V.}}:
\byear{2009},
\batitle{{Enhancement of the guide field during the current sheet formation in
  the three-dimensional magnetic configuration with an X line}}.
\bjtitle{Phys. Lett. A}
\bvolume{373},
\bfpage{1460}.
\doiurl{https://doi.org/10.1016/j.physleta.2009.02.037}.
\adsurl{2009PhLA..373.1460F}.
\end{barticle}
\endbibitem

\bibitem[\protect\citeauthoryear{{Furth}, {Killeen}, and
  {Rosenbluth}}{1963}]{1963PhFl....6..459F}
\begin{barticle}
\bauthor{\bsnm{{Furth}}, \binits{H.P.}},
\bauthor{\bsnm{{Killeen}}, \binits{J.}},
\bauthor{\bsnm{{Rosenbluth}}, \binits{M.N.}}:
\byear{1963},
\batitle{{Finite-Resistivity Instabilities of a Sheet Pinch}}.
\bjtitle{\physf}
\bvolume{6},
\bfpage{459}.
\doiurl{https://doi.org/10.1063/1.1706761}.
\adsurl{1963PhFl....6..459F}.
\end{barticle}
\endbibitem

\bibitem[\protect\citeauthoryear{{Gorbachev} and
  {Somov}}{1988}]{1988SoPh..117...77G}
\begin{barticle}
\bauthor{\bsnm{{Gorbachev}}, \binits{V.S.}},
\bauthor{\bsnm{{Somov}}, \binits{B.V.}}:
\byear{1988},
\batitle{{Photospheric Vortex Flows as a Cause for Two-Ribbon Flares - a
  Topological Model}}.
\bjtitle{\solphys}
\bvolume{117},
\bfpage{77}.
\doiurl{https://doi.org/10.1007/BF00148574}.
\adsurl{1988SoPh..117...77G}.
\end{barticle}
\endbibitem

\bibitem[\protect\citeauthoryear{{Gorbachev}
  et~al.}{1988}]{1988SvA....32..308G}
\begin{barticle}
\bauthor{\bsnm{{Gorbachev}}, \binits{V.S.}},
\bauthor{\bsnm{{Kelner}}, \binits{S.R.}},
\bauthor{\bsnm{{Somov}}, \binits{B.V.}},
\bauthor{\bsnm{{Shvarts}}, \binits{A.S.}}:
\byear{1988},
\batitle{{A New Topological Approach to the Question of the Trigger for Solar
  Flares}}.
\bjtitle{\sovast}
\bvolume{32},
\bfpage{308}.
\adsurl{1988SvA....32..308G}.
\end{barticle}
\endbibitem

\bibitem[\protect\citeauthoryear{{Ledentsov}}{2021}]{2021SoPh..296...74L}
\begin{barticle}
\bauthor{\bsnm{{Ledentsov}}, \binits{L.}}:
\byear{2021},
\batitle{{Thermal Trigger for Solar Flares I: Fragmentation of the Preflare
  Current Layer}}.
\bjtitle{\solphys}
\bvolume{296},
\bfpage{74}.
\doiurl{https://doi.org/10.1007/s11207-021-01817-1}.
\adsurl{2021SoPh..296...74L}.
\end{barticle}
\endbibitem

\bibitem[\protect\citeauthoryear{{Ledentsov} and
  {Somov}}{2015}]{2015PhyU...58..107L}
\begin{barticle}
\bauthor{\bsnm{{Ledentsov}}, \binits{L.S.}},
\bauthor{\bsnm{{Somov}}, \binits{B.V.}}:
\byear{2015},
\batitle{{Discontinuous plasma flows in magnetohydrodynamics and in the physics
  of magnetic reconnection}}.
\bjtitle{Phys. Uspekhi}
\bvolume{58},
\bfpage{107}.
\doiurl{https://doi.org/10.3367/UFNe.0185.201502a.0113}.
\adsurl{2015PhyU...58..107L}.
\end{barticle}
\endbibitem

\bibitem[\protect\citeauthoryear{{Oreshina}, {Oreshina}, and
  {Somov}}{2012}]{2012A&A...538A.138O}
\begin{barticle}
\bauthor{\bsnm{{Oreshina}}, \binits{A.V.}},
\bauthor{\bsnm{{Oreshina}}, \binits{I.V.}},
\bauthor{\bsnm{{Somov}}, \binits{B.V.}}:
\byear{2012},
\batitle{{Magnetic-topology evolution in NOAA AR 10501 on 2003 November 18}}.
\bjtitle{\aap}
\bvolume{538},
\bfpage{A138}.
\doiurl{https://doi.org/10.1051/0004-6361/201117620}.
\adsurl{2012A&A...538A.138O}.
\end{barticle}
\endbibitem

\bibitem[\protect\citeauthoryear{{Priest} and
  {Forbes}}{2002}]{2002A&ARv..10..313P}
\begin{barticle}
\bauthor{\bsnm{{Priest}}, \binits{E.R.}},
\bauthor{\bsnm{{Forbes}}, \binits{T.G.}}:
\byear{2002},
\batitle{{The magnetic nature of solar flares}}.
\bjtitle{\aapr}
\bvolume{10},
\bfpage{313}.
\doiurl{https://doi.org/10.1007/s001590100013}.
\adsurl{2002A&ARv..10..313P}.
\end{barticle}
\endbibitem

\bibitem[\protect\citeauthoryear{{Shibata} and
  {Magara}}{2011}]{2011LRSP....8....6S}
\begin{barticle}
\bauthor{\bsnm{{Shibata}}, \binits{K.}},
\bauthor{\bsnm{{Magara}}, \binits{T.}}:
\byear{2011},
\batitle{{Solar Flares: Magnetohydrodynamic Processes}}.
\bjtitle{\lrsf}
\bvolume{8},
\bfpage{6}.
\doiurl{https://doi.org/10.12942/lrsp-2011-6}.
\adsurl{2011LRSP....8....6S}.
\end{barticle}
\endbibitem

\bibitem[\protect\citeauthoryear{{Somov}}{2012}]{2006ASSL..340.....S}
\begin{bbook}
\bauthor{\bsnm{{Somov}}, \binits{B.V.}}:
\byear{2012},
\bbtitle{{Plasma Astrophysics. Part I: Fundamentals and Practice. Second
  Edition}}
\bseriesno{391},
\bpublisher{Astrophys. Space Sci. Library, ASSL}.
\doiurl{https://doi.org/10.1007/978-1-4614-4283-7}.
\adsurl{2012ASSL..391.....S}.
\end{bbook}
\endbibitem

\bibitem[\protect\citeauthoryear{{Somov}}{2013}]{2006ASSL..341.....S}
\begin{bbook}
\bauthor{\bsnm{{Somov}}, \binits{B.V.}}:
\byear{2013},
\bbtitle{{Plasma Astrophysics. Part II: Reconnection and Flares. Second
  Edition}}
\bseriesno{392},
\bpublisher{Astrophys. Space Sci. Library, ASSL}.
\doiurl{https://doi.org/10.1007/978-1-4614-4295-0}.
\adsurl{2013ASSL..392.....S}.
\end{bbook}
\endbibitem

\bibitem[\protect\citeauthoryear{{Somov} and
  {Titov}}{1985a}]{1985SoPh...95..141S}
\begin{barticle}
\bauthor{\bsnm{{Somov}}, \binits{B.V.}},
\bauthor{\bsnm{{Titov}}, \binits{V.S.}}:
\byear{1985}a,
\batitle{{Magnetic Reconnection in a High Temperature Plasma of Solar Flares}}.
\bjtitle{\solphys}
\bvolume{95},
\bfpage{141}.
\doiurl{https://doi.org/10.1007/BF00162642}.
\adsurl{1985SoPh...95..141S}.
\end{barticle}
\endbibitem

\bibitem[\protect\citeauthoryear{{Somov} and
  {Titov}}{1985b}]{1985SoPh..102...79S}
\begin{barticle}
\bauthor{\bsnm{{Somov}}, \binits{B.V.}},
\bauthor{\bsnm{{Titov}}, \binits{V.S.}}:
\byear{1985}b,
\batitle{{Magnetic Reconnection in a High-Temperature Plasma of Solar Flares -
  Part Two - Effects Caused by Transverse and Longitudinal Magnetic Fields}}.
\bjtitle{\solphys}
\bvolume{102},
\bfpage{79}.
\doiurl{https://doi.org/10.1007/BF00154039}.
\adsurl{1985SoPh..102...79S}.
\end{barticle}
\endbibitem

\bibitem[\protect\citeauthoryear{{Somov} and
  {Verneta}}{1993}]{1993SSRv...65..253S}
\begin{barticle}
\bauthor{\bsnm{{Somov}}, \binits{B.V.}},
\bauthor{\bsnm{{Verneta}}, \binits{A.I.}}:
\byear{1993},
\batitle{{Tearing Instability of Reconnecting current Sheets in Space
  Plasmas}}.
\bjtitle{\ssr}
\bvolume{65},
\bfpage{253}.
\doiurl{https://doi.org/10.1007/BF00754510}.
\adsurl{1993SSRv...65..253S}.
\end{barticle}
\endbibitem

\bibitem[\protect\citeauthoryear{{Syrovatskii}}{1956}]{1956TrFIAN...13..64S}
\begin{barticle}
\bauthor{\bsnm{{Syrovatskii}}, \binits{S.I.}}:
\byear{1956},
\batitle{{Some properties of discontinuity surfaces in magnetohydrodynamics}}.
\bjtitle{Tr. Fiz. Inst. im. P.N. Lebedeva, Akad. Nauk SSSR {\rm [in Russian]}}
\bvolume{8},
\bfpage{13}.
\end{barticle}
\endbibitem

\bibitem[\protect\citeauthoryear{{Syrovatskii}}{1958}]{1958ForPh...6..437S}
\begin{barticle}
\bauthor{\bsnm{{Syrovatskii}}, \binits{S.I.}}:
\byear{1958},
\batitle{{Magnetohydrodynamik}}.
\bjtitle{Fortschritte der Physik}
\bvolume{6},
\bfpage{437}.
\doiurl{https://doi.org/10.1002/prop.19580060902}.
\adsurl{1958ForPh...6..437S}.
\end{barticle}
\endbibitem

\bibitem[\protect\citeauthoryear{{Syrovatskii}}{1971}]{1971JETP...33..933S}
\begin{barticle}
\bauthor{\bsnm{{Syrovatskii}}, \binits{S.I.}}:
\byear{1971},
\batitle{{Formation of Current Sheets in a Plasma with a Frozen-in Strong
  Magnetic Field}}.
\bjtitle{Soviet Journal of Experimental and Theoretical Physics}
\bvolume{33},
\bfpage{933}.
\adsurl{1971JETP...33..933S}.
\end{barticle}
\endbibitem

\bibitem[\protect\citeauthoryear{{Verneta} and
  {Somov}}{1993}]{1993ARep...37..282V}
\begin{barticle}
\bauthor{\bsnm{{Verneta}}, \binits{A.I.}},
\bauthor{\bsnm{{Somov}}, \binits{B.V.}}:
\byear{1993},
\batitle{{Effect of compressibility on the development of tearing instability
  in a non-neutral current sheet in the solar atmosphere}}.
\bjtitle{Astron. Rep.}
\bvolume{37},
\bfpage{282}.
\adsurl{1993ARep...37..282V}.
\end{barticle}
\endbibitem

\end{thebibliography}

\end{article} 

\end{document}